\documentclass[prb,twocolumn,showpacs,floatfix]{revtex4}

\usepackage{latexsym,amsmath,graphics,graphicx}
\usepackage [english] {babel}

\begin{document}
\title{Vacancy complexes with oversized impurities in Si and Ge}
\author{H.~H\"ohler, N.~Atodiresei, K.~Schroeder,
R.~Zeller, and P.~H.~Dederichs}
\affiliation{Institut f\"ur
Festk\"orperforschung, Forschungszentrum J\"ulich, D-52425 J\"ulich,
Germany}
\date{\today}

\begin{abstract}
In this paper we examine the electronic and geometrical structure
of impurity-vacancy complexes in Si and Ge. Already Watkins suggested
that in Si the pairing of Sn with the vacancy 
produces a complex with the Sn-atom
at the bond center and the vacancy split into two half vacancies on
the neighboring sites. Within the framework of density-functional
theory we use two complementary {\it{ab initio}} methods, the pseudopotential
plane wave (PPW) method and the all-electron Kohn-Korringa-Rostoker (KKR)
method, to investigate the structure of vacancy complexes with 11 different
sp-impurities. For the case of Sn in Si, we confirm the split configuration
and obtain good agreement with EPR data of Watkins. In general we find 
that all impurities of the 5sp and 6sp series in Si and Ge prefer
the split-vacancy configuration, with an energy gain of 0.5 to 1 eV
compared to the substitutional complex. On the other hand, impurities
of the 3sp and 4sp series form a (slightly distorted) substitutional
complex. Al impurities show an exception from this rule, forming a
split complex in Si and a strongly distorted substitutional complex in Ge.
We find a strong correlation of these data with the size of the isolated
impurities, being defined via the lattice relaxations of the
nearest neighbors.
\end{abstract}


\maketitle

\section{Introduction}
\label{secI}
Intrinsic defects and their complexes with impurities play an
important role in semiconductor physics. In this paper we present
{\it{ab initio}} calculations for vacancy-impurity complexes in Si and Ge.
Based on EPR measurements, Watkins\cite{Watkins:75.1} has already
shown that in Si the Sn-vacancy complex prefers an exotic
configuration, i.e.\ a split-vacancy complex with the Sn atom on the
bond-center position and the vacancy split into two half-vacancies on
the nearest neighbor (NN) sites.  This configuration has been
supported by ab initio calculations for Sn in Si by Nylandsted Larsen
et al.\cite{Kaukonen:00.1} and Kaukonen et al.\cite{Kaukonen:01.1},
reporting a small energy preference of 0.045 eV for this complex
compared to the configuration with substitutional Sn and the vacancy
on NN sites.

Recently we have studied by density functional calculations the
Cd-vacancy and Cd-interstitial complexes in Si and
Ge\cite{Hoehler:04.1}, aiming at understanding the electric field
gradients (EFG) measured in perturbed angular correlation experiments
for the ${}^{111}\mathrm{In}/{}^{111}\mathrm{Cd}$ probe atom.  We find
that both in Si and Ge the substitutional Cd-vacancy complex is
instable and relaxes into the highly symmetrical Cd-split-vacancy
complex, being about 1eV lower in energy than the substitutional
configuration. In the split configuration the Cd atom hybridizes very
weakly with the six Si or Ge nearest neighbors, resulting in a nearly
isotropic charge density and an extremely small EFG. In this way we
find good agreement with PAC measurements and can uniquely assign the
very small measured Cd EFG's of 28 MHz in Si and 54 MHz in Ge to the
Cd-split-vacancy complex.  In parallel to our calculations Alippi et.
al. show in a recent publication\cite{Alippi:04.1}, that in Si also In
impurities form such a split configuration with the vacancy,
exhibiting an unusually large binding energy of 2.4 eV.

The present paper has a twofold aim. First we present {\it{ab initio}}
calculations for the electronic and geometrical structure of the
Sn-vacancy complex in Si and Ge. In particular we concentrate on the
hyperfine properties of this defect for the different charge states,
i.e.\ the hyperfine fields, the isomer shifts and the electric field
gradients, and compare with available experimental data by
Watkins\cite{Watkins:75.1} and others\cite{Weyer:80.1,Weyer:80.2}.
These properties have not been calculated so far, since they are
difficult to 
obtain by the standard pseudopotential plane wave method.
In the second part of the paper the structure of impurity-vacancy
complexes for other impurities is calculated, in particular heavy
impurities with larger sizes than Si or Ge atoms. For the elements Cd,
In, Sn and Sb of the 5sp-series we find that both in Si and Ge the
split-vacancy impurity complex is preferred over the substitutional
one by energies between 0.5 to 1 eV, being slightly larger in Si than
in Ge. The results for the even heavier element Bi of the 6sp series
suggest, that in Si and Ge this is the stable configuration for all
oversized impurities.  In order to discuss the importance of impurity
size for the relative stability of the substitutional and split
configurations we calculate the lattice relaxations of the nearest
neighbors for a large series of isolated impurities from the 3sp, 4sp,
and 5sp series. In general, we find a good qualitative correlation of
the sign and size of the NN relaxations with the stability of the two
configurations, although a strict one-to-one correspondence is not
valid.

\section{Theoretical Methods}
\label{secII}
All calculations are based on density functional theory in the local
density approximation\cite{Vosko:80.1}. Two different methods have
been used to solve the Kohn-Sham equations.  The first one is the
pseudopotential plane wave (PPW) method, which has mostly been used to
investigate the configuration and stability of the impurity-vacancy
configurations. These configurations, as well as the relaxations of
the neighboring atoms, have been recalculated by the KKR-Green-function
method\cite{Brasp:84.1,Nikos:02.1} for the Sn-impurity, which as an
all-electron method allows also to calculate electric field gradients,
isomer shifts and hyperfine fields\cite{Akai:90.1,Blaha:88.1}.

The PPW method approximates the inhomogeneous systems containing defect 
complexes by periodically arranged large supercells,
and uses plane waves to expand the electronic wave functions. 
This has the advantage that band-structure methods can be used to determine 
the electronic structure, and total energies and forces on the atoms can be calculated
without difficulty for arbitrary arrangements of the atoms in the supercell.
We have used norm-conserving Kleinman-Bylander
(KB)-pseudopotentials\cite{K-B:82} for all atoms considered, 
where the $s,p$-valence electrons are treated by projectors,
and the $d(l=2)$-component is used as a local potential. 
Our {\tt{EStCoMPP}}-program\cite{Berger:00} is fully parallelized and can efficiently
handle supercells with up to 300 atoms.
For the calculations of impurity-vacancy complexes we used a (111)-oriented supercell 
with the basis vectors
$\vec{b}_1 = 3 a (0, -1, 1); \vec{b}_2 = 3 a (-1, 1, 0); \vec{b}_3 = 2 a (1, 1, 1)$
containing 108 atoms. $a$ is the theoretical lattice constant ( 10.71$a_B$ for Ge
with $E_{cut} =$ 11.56 Ry,
and 10.25$a_B$ for Si with $E_{cut} =$ 9 Ry; Bohr radius $a_B$ = 0.529177 \AA). 
The impurity-complexes were placed in the middle of the cell. 
We used $C_{3v}$ symmetry explicitly for all configurations.
The isolated substitutional impurities were calculated with a 
2$\times$2$\times$2 $a^3$ cubic cell 
containing 64 atoms. The impurities were located at the central site, 
and the T$_d$ site-symmetry is enforced.
A plane-wave basis set equivalent to $6\times6\times6$ 
Monkhorst-Pack\cite{Monkhorst-Pack:77} $\vec{k}$-points was used
which yields three inequivalent $\vec{k}$-points in the irreducible part of the 
Brillouin-zone for the 108-atom-supercell and five inequivalent 
$\vec{k}$-points for the 64-atom-supercell. 
We used a plane-wave cut-off $E_{cut} =$ 9 Ry for impurities in Si, 
and $E_{cut} =$ 11.56 Ry for impurities in Ge. 
The atoms belonging to the impurity-complexes and their nearest neighbors 
were relaxed until the forces on all atoms were less than 0.1 mRy/$a_B$. 
We checked that forces on further neighbors (which were not moved) were 
less than 20 mRy/$a_B$.
The Sn-complexes in both hosts were recalculated for a cut-off energy of 13.67 Ry, 
and we found no significant changes of the configurations or the energy differences 
for the tested configurations. We estimate that the positions of all relaxed 
atoms are determined with an accuracy better than 0.01 $a_B$, and the energy 
differences are accurate to about 0.1 eV.

In the KKR-Green-function method the calculation is divided in two steps. First the Green function of the host is determined. In a second step the host Green function is used to determine the Green function of the crystal with a single impurity by a Dyson equation. For details we refer to reference\cite{Nikos:97.1}. All calculations are performed with an angular momentum cut-off of $l_{max}=4$. For the host crystal we use the LDA lattice constants\cite{Settels:99.3} of Si (10.21$a_0$) and Ge (10.53$a_0$). A k-mesh of $30\times 30\times 30$ $\vec{k}$-points in the full Brillouin zone is used. We use the screened KKR-formalism\cite{Zeller:95.1,Zeller:97.1,Nikos:02.1}
with the TB-structure constants determined from a cluster of 65 repulsive potentials of 4 Ry heights. The diamond structure is described by a unit cell with 4 basis sites, two for host atoms and two for vacant sites. For the Green function of the defective systems, we allow 77 potentials of the defect and the surrounding host atoms to be perturbed, which are then calculated selfconsistently with proper embedding into the host crystal. All calculations include the fully anisotropic potentials in each cell and thus allow the reliable calculation of forces, lattice relaxations and electric field gradients.
The  Coulomb potential can be expressed in terms of $n(\vec{r})$ and therefore the force $\vec{F}^n$ on atom $n$ can be derived with the use of the ``ionic'' Hellmann-Feynman theorem\cite{Nikos:97.1}.
The hyperfine parameters of interest in the present paper are the isomer shifts (IS), the Fermi contact term of the hyperfine field (HF) and the quadrupole splitting ($\Delta^*$).
The last quantity is determined by the electric field  at the probe-atom site.
In a non-relativistic treatment the isomer shift can be calculated from the
charge density $n(0)$ at the nucleus
\begin{eqnarray}
\label{alpha}
     IS          & = &\alpha \ {n(0)},
\end{eqnarray}
where $\alpha$ is the calibration constant. The hyperfine field is given by
the Fermi contact contribution in terms of the magnetization density
$m(0)$ at the nucleus.
\begin{eqnarray}
\Delta{HF}= {\frac{8\pi}{3}}\mu_B \ m(0),
\end{eqnarray}
Finally the quadrupole splitting is calculated from
\begin{eqnarray}
 \Delta^*    & = &
               {\frac{1}{2}}{\frac{e\vert QV_{zz}\vert}{E_{source}}}\times c,
\end{eqnarray}
where $Q$ is the quadrupole moment of the 
probe-nucleus and $V_{zz}$
is the electric field gradient along the main (111) axis.
The parameter $c$ is the speed of light and $E_{source}$ is the energy 
of the emitted $\gamma$-ray, which is 23.8 keV for Sn.
The tensor of the EFG is given by the second derivatives of the Coulomb
potential and can be written as\cite{Blaha:88.1}:
\begin{eqnarray}
\label{defefg2}
{\bf V}_{\alpha\beta} & = & \sum_{m=-2}^2 \tilde{V}_{2,m}(0)\partial_{\alpha}\partial_{\beta}(r^2Y_{2m}(\vec{r})),\\
\nonumber \tilde{V}_{2,m}(0) & = & I_1 \ + \ I_2,\\
\nonumber I_1 & = & \frac{8\pi}{5}\int_0^{R_{MT}} {r'}^2 \frac{n_{2m}(r')}{{r'}^3}dr'; \\
\nonumber I_2 & = & \frac{V_{2m}(R_{MT})}{{R_{MT}}^2}-\frac{8\pi}{5}\int_0^{R_{MT}} \frac{n_{2m}(r'){r'}^4}{{R_{MT}}^5}dr',
\end{eqnarray} 
where $Y_{2m}(\vec{r})$ are spherical harmonics, $\tilde{V}_{2,m}(r)$ are the $l=2$ components of the Coulomb potential and $R_{MT}$ is the muffin tin radius of the probe atom. Furthermore $n_{2m}(r)$ are 
the $l=2$ components ($m=-2,\dots,+2$) of the radial charge density.
We note that just the $l=2$ components are needed for the
EFGs, while the $l=1$ components enter into the forces.
Consequently, the EFG vanishes for sites with cubic or tetrahedral symmetry. For the cell division of the crystal we used a generalized Voronoi construction.
The impurity cell at the bond center was constructed such that it is slightly larger than the cells of the host atoms, thus avoiding to decrease the muffin tin radius of real atoms by more than 16\%.
For the treatment of occupied states in the band gap, we have used a separate energy contour for gap energies. Since we used group theory we 
were 
able to occupy these gap-states according to their symmetry, in this way avoiding to calculate the wave functions of the localized gap states. 
   
\section{Electronic Structure of S\lowercase{n}-Vacancy Complexes}
\label{secIII}
\begin{figure}[t]
\begin{center}
\includegraphics[scale=0.6]{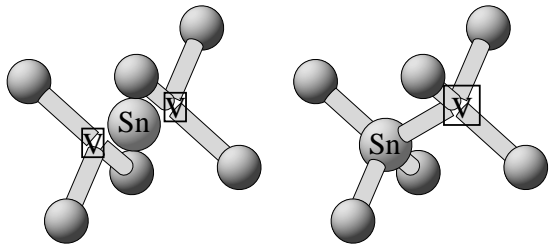}
\caption{The Sn-split-vacancy configuration (left) and the substitutional Sn-vacancy configuration (right).}
\label{fig:1}
\end{center}
\end{figure}

As predicted already many years ago by Watkins\cite{Watkins:75.1} on the basis of EPR measurements, and as has also be confirmed by {\it{ab initio}} calculations\cite{Kaukonen:00.1,Kaukonen:01.1},
the stable configuration of the Sn-vacancy pair in Si consists of a Sn-atom at the bond-center position and the vacancy split in two halves on the two (empty) nearest neighbor positions.
This split-vacancy configuration is shown schematically in Fig. \ref{fig:1} together with the ``normal'' substitutional complex.
As we have found recently\cite{Hoehler:04.1} the same split-vacancy configuration occurs also upon pairing with Cd-atoms, yielding both in Si and Ge very small electric field gradients, which allow to identify this complex uniquely.
In this configuration the Sn or the Cd atom are coordinated by six host atoms at the relatively large distance of 1.26 times the nearest neighbor distance (in the unrelaxed configuration).

\begin{figure}[t]
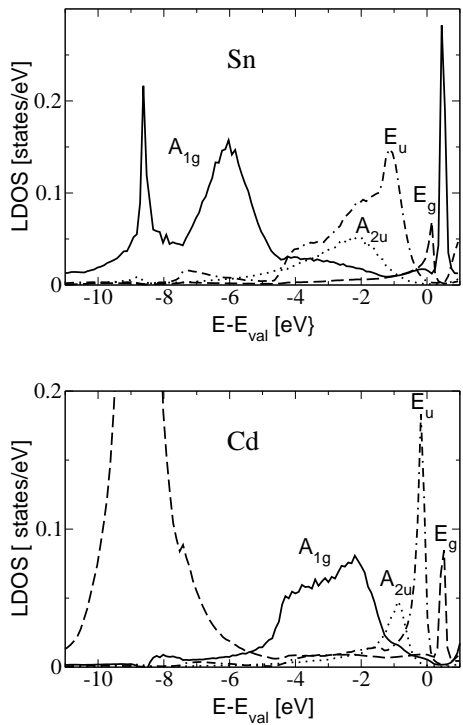

\center{\includegraphics*[scale=0.25]{dos_sn_in_ge_sym.eps}\\[0.5cm]
\includegraphics*[scale=0.25]{cd_in_ge_sym_snpaper.eps}
\caption{LDOS projected on the irreducible subspaces of the $D_{3d}$ group
at the Sn site (above) and at the Cd site (below) for
the impurity split-vacancy complex in Ge.
The nearest neighbors are fixed on ideal lattice positions.\label{fig:2}}}
\end{figure}

The local density of states (LDOS) of this complex at the Sn-site in the Ge host is shown in Fig. \ref{fig:2} (upper panel), together with the corresponding LDOS for the Cd split-vacancy complex in Ge in Fig. \ref{fig:2} (lower panel). In the calculation the Fermi level is fixed at the maximum $E_{val}$ of the valence band. The different peaks are labeled by the irreducible subspaces $A_{1g}$, $A_{2u}$, $E_{u}$ and $E_{g}$ of the $D_{3d}$ group, representing the point symmetry of the complex.
Compared to the Cd complex with the $d$-level at about -9 eV,
the $d$-level of Sn is fully localized and at much lower energies. However otherwise the level scheme is very similar: the $A_{1g}$ state at relative low energies, the $A_{2u}$ and the doubly degenerate $E_u$ state slightly below the Fermi level, and the doubly degenerate $E_g$ state in the gap and, in the case of Sn, a second $A_{1g}$ state at higher energies.
Since the occupied $A_{1g}$, $A_{2u}$ and $E_u$ states accommodate 8 electrons,
of which the 6 neighboring host atoms provide 6 (one dangling bond each), the vacancy complex 
with Cd (with 2 valence electrons) is neutral for $E_F=E_{val}$ ($[{\mathrm{CdV}}]^0$),
 and the complex with Sn (with 4 valence electrons) is doubly positively charged 
($[{\mathrm{SnV}}]^{2+}$).
As discussed in reference\cite{Hoehler:04.1}, the level sequence is basically determined by the divacancy.
The Cd and Sn-atoms can be considered as ionized Cd$^{2+}$ and Sn$^{4+}$ ions inserted in the center of the divacancy and only weakly hybridizing with the 6 nearest neighbors.
The attractive ionic potentials shift the divacancy states to lower energies,
in particular the fully symmetrical $A_{1g}$ state,
being the only state affected in the first Born approximation with
respect to the potential.
This effect is naturally stronger for the Sn$^{4+}$ ion than for the Cd$^{2+}$ ion, but the LDOS are very similar.
Of course, also Cd and Sn s and p states, localized at energies above $E_{val}$, are hybridized into the occupied states, so that the local charge in the impurity cell is about 1 electron in the case of Cd in Ge and 2 electrons in the case of Sn in Ge. In fact, the $A_{1g}$ state above $E_F$ can be considered as the genuine Sn 5s states, which is, however, not occupied.

By occupying the $E_g$ state in the gap, we obtain complexes with additional electron charges, i.e.\ $[{\mathrm{SnV}}]^{+}$, $[{\mathrm{SnV}}]^{0}$, ... and analogously $[{\mathrm{CdV}}]^{-}$, $[{\mathrm{CdV}}]^{2-}$, ... complexes.
Both the $[{\mathrm{SnV}}]^{+}$ and the $[{\mathrm{SnV}}]^{0}$ complexes are magnetic with total moments of 1 and 2 $\mu_B$, and with small local moments of 0.045 $\mu_B$ (0.008 $\mu_B$) and 0.091 $\mu_B$ (0.016 $\mu_B$) on the Sn site in Si (Ge).
Although we have calculated only the three charge states $[{\mathrm{SnV}}]^{2+}$, $[{\mathrm{SnV}}]^{+}$ and $[{\mathrm{SnV}}]^{0}$, calculations as well as experiments show that also the negatively charged $[{\mathrm{SnV}}]^{-}$ and $[{\mathrm{SnV}}]^{2-}$states exist.
Since for the complex $[{\mathrm{SnV}}]^{0}$ the $E_g$ majority state is completely filled
having a moment of 2 $\mu_B$, we expect that also the $[{\mathrm{SnV}}]^{-}$ state is
magnetic with a total moment of 1 $\mu_B$ and local moments similar to the $[{\mathrm{SnV}}]^{+}$ complex, while the $[{\mathrm{SnV}}]^{2-}$ is non-magnetic. The existence of these five charge states arises from the fact, that the $E_g$ state is very extended and has moreover a nodal plane at the Sn-site so that Coulomb effects are very small. This also explains the very small local moments. The situation is quite similar for Cd complexes, except that the local moments are somewhat smaller in Si, e.g. 0.024 $\mu_B$ for $[{\mathrm{CdV}}]^{-}$. The same level sequences we expect also for other 5sp impurities to occur. For instance for In, the charge state should be $[{\mathrm{InV}}]^+$, if $E_F$ is fixed at $E_{val}$, and the $[{\mathrm{InV}}]^{0}$, $[{\mathrm{InV}}]^{-}$ and $[{\mathrm{InV}}]^{2-}$  configurations should be magnetic. In all calculations the complexes in Si are very similar to the above results for the Ge hosts. 

The calculations yield sizeable relaxations of the neighboring host atoms towards the Sn atoms, which are practically radially symmetrical and increase with the charge state. For $[{\mathrm{SnV}}]^{2+}$ the NN Ge atoms relax by $2.9\%$ of the NN distance, for $[{\mathrm{SnV}}]^{+}$ by $4.2\%$ and $[{\mathrm{SnV}}]^{0}$ by $5.4\%$. In Si the corresponding relaxations are somewhat larger: $6\%$, $6.2\%$ and $6.4\%$.
The relaxations arise from the attraction of the neighboring electrons
to the Sn$^{4+}$ ion and increase with occupation of the $E_g$ gap states,
since these states are dangling-bond like and mostly localized
at the host atoms.
Therefore the neutral $[{\mathrm{SnV}}]^0$ state has larger relaxations than the $[{\mathrm{SnV}}]^{2+}$, which is opposite to the normal expectations.      

\section{Hyperfine properties of S\lowercase{n}-vacancy complexes}
\label{secIV}

\begin{table}[t]
\center{
\begin{tabular}{lcc}
\hline
\hline
Host &       Si         & Ge\\
        &  $H_{HF}$ [kG]              & $H_{HF}$ [kG]\\
\hline
$\rm{[V|Sn|V]^{+}}$ & 139 & 6.00 \\
$\rm{[V|Sn|V]^{0}}$ & 211 & 14.59 \\
EXP           & 241 (386MHz)& \\
\hline
\hline
\end{tabular}\\[0.25cm]
\begin{tabular}{lcc}
\hline
\hline
Host &       Si         & Ge\\
        &  $H_{HF}$ [kG]              & $H_{HF}$ [kG]\\
\hline
$\rm{[V|Sn|V]^{+}}$ & -39.59 & -44.46\\
$\rm{[V|Sn|V]^{0}}$ & -82.96 &  -94.19 \\
EXP           &  -91.02 (-77.00MHz) & \\
\hline
\hline
\end{tabular}}
\caption{The calculated hyperfine field (HF) at the Sn atom (above) and
the nearest neighbors (below) for the Sn-split-vacancy in Si and Ge is shown. A comparison to the experimental value of Watkins\cite{Watkins:75.1} is given. \label{table:1}}
\end{table}
Already Watkins\cite{Watkins:75.1} has measured by EPR hyperfine fields at the neighboring Si atoms adjacent to a ${\mathrm{SnV}}$ complex, which he identified as the neutral $[{\mathrm{SnV}}]^0$ complex with total spin $S=1$. Table \ref{table:1} shows the calculated hyperfine fields at the Sn atom and the six Si or Ge neighbors adjacent to the magnetic $[{\mathrm{SnV}}]^+$ and $[{\mathrm{SnV}}]^0$ split complexes 
and a comparison with the experimental values\cite{Watkins:75.1,Fanciulli:00.1}.
First we note, that the field for $[{\mathrm{SnV}}]^0$ is
about twice the field of $[{\mathrm{SnV}}]^+$.
This is a consequence of the large extent of the $E_g$ state,
representing locally a small perturbation, so that the forces increase
linearly with the occupancy.
Second, for $[{\mathrm{SnV}}]^0$ we obtain good agreement with the experimental values.
Here we note, that our calculations are non-relativistic, so that the relativistic enhancement\cite{Akai:90.1} of the hyperfine field is missing, which would presumably improve the agreement. For the $[{\mathrm{SnV}}]^-$ complex we expect a similar hyperfine field as for the $[{\mathrm{SnV}}]^+$ complex, since the total moment is the same. We have chosen in our calculations the spin-polarization such, that the total moment has a negative sign (spin down occupation).
As a consequence the local moments for all sites are negative,
since the $E_g$ state determines the magnetic properties alone.
That means that $m_{loc}/H_{HF}<0$ at the Sn-site and $m_{loc}/H_{HF}>0$ at the Si nearest neighbor site.
The change of sign of the hyperfine field from Si to Sn can be explained
due to the different exchange interaction mechanism between the d-like
``spin down'' $E_g$ electrons and the ``spin down'' and ``spin up''
s-core-electrons at the Sn-site. In the first case the interaction is
attractive, leading to a smaller s charge density at the nucleus,
whereas in the latter case the situation is opposite.
In total the hyperfine field changes its sign in comparison to the Si site.
In addition to the hyperfine fields we have
calculated the isomer shifts and quadrupole splittings. These
observables can be measured by the M\"ossbauer spectroscopy.
Measurements using the$\phantom{x}^{119}\mathrm{Sn}$ isotope in Si and
Ge were performed by Weyer et. al.\cite{Weyer:80.1,Weyer:80.2} in 1980,
which however do not agree with our calculated results.
From various 
private
discussions we conclude, that in these early experiments
clustering of Sn-impurities might have occurred, leading to more
complicated structures. On the other hand, our results are in good
agreement with unpublished measurements of
R. Sielemann (private communication).

In Table {\ref{isosnsi}} the results for Si and in Table {\ref{isosnge}}
the results for Ge are presented. The calibration constant $\alpha$ was
determined by calculating $n(0)$ for several Sn-defects, for which
well known isomer shifts exist.
Svane et. al.\cite{Svane:88.1,Svane:97.1} already
calculated  in a relativistic treatment an $\alpha$ of
 0.092 $a_0^3$ ${\mathrm{mms}}^{-1}$ , which is in good agreement to ours with taking the Shirley factor\cite{Shirley:64.1}
 for Sn of 2.48 into account. A closer view
on the isomer shifts for the Sn-split-vacancy configuration
(in relation to the M\"ossbauer line of $\alpha$-Sn) 
 shows a decreasing charge density at the
nucleus with the occupation of the $E_g$ state in the band gap. 
This behavior can be explained due to the screening impact of the d-like
$E_g$-electrons at the Sn-site on the s-electrons responsible for the IS.
As in the Cd case ($Q_{Cd}=0.83$ ) the Sn-EFG's are rather small,
since hybridization with
the neighbors is weak. It is obvious, that the EFG increases with occupying 
the $E_g$ state, since $p_x$ and $p_y$ charges give positive
contributions to the EFG at the Sn-site.  
    
\begin{table}[t]
\centerline{
\begin{tabular}{lcccc}
\hline
\hline
system & $n(\vec{r}=0)$      & $IS$               & $EFG$   &  $ \Delta^*$  \\
       & $[a_0^{-3}]$ & $[{\mathrm{mms}}^{-1}]$ & [MHz] & $[{\mathrm{mms}}^{-1}]$ \\
\hline
$\alpha$-Sn & 87759.82 & ref & 0.00 & 0.00 \\
subst. Sn   & 87758.81 & -0.2200 & 0.00 & 0.00  \\       
$\rm{[V|Sn|V]^{2+}}$ & 87759.71 & -0.0257 &  -14.90 & 0.3875   \\
$\rm{[V|Sn|V]^{+}}$  & 87759.36 & -0.1076 &  -10.59 & 0.2713\\
$\rm{[V|Sn|V]^{0}}$ & 87758.99 & -0.1942 &  -5.62 & 0.1473\\
\hline
\hline
\end{tabular}}
\caption{Isomer shifts and quadrupole splitting for
  the Sn-vacancy complexes in Si. ($\alpha=0.2178
  \hspace{0.15cm}a_0^3 \hspace{0.15cm}{\mathrm{mms}}^{-1}$, $E_{source}=23.8$ keV, $Q=-0.124$ barn)
\label{isosnsi}}
\centerline{
\begin{tabular}{lcccc}
\hline
\hline
system & $n(\vec{r}=0)$      & $IS$               & $EFG$   &  $ \Delta^*$  \\
       & $[a_0^{-3}]$ & $[{\mathrm{mms}}^{-1}]$ & [MHz] & $[{\mathrm{mms}}^{-1}]$ \\
\hline
$\alpha$-Sn & 87759.82 & ref & 0.00 & 0.00 \\
subst. Sn   & 87759.26 & -0.1400 & 0.00 & 0.00  \\       
$\rm{[V|Sn|V]^{2+}}$ & 87760.61 & 0.17 & -12.05 & 0.31   \\
$\rm{[V|Sn|V]^{+}}$  & 87760.55 & 0.16 & -8.08 & 0.1938\\
$\rm{[V|Sn|V]^{0}}$  & 87760.46 &  0.14 & -3.97 & 0.1027\\
\hline
\hline
\end{tabular}}
\caption{Isomer shifts and quadrupole splitting for
  the Sn-vacancy complexes in Ge. ($\alpha=0.2178
  \hspace{0.15cm}a_0^3 \hspace{0.15cm}{\mathrm{mms}}^{-1}$, $E_{source}=23.8$ keV, $Q=-0.124$ barn)
\label{isosnge}}
\end{table}

\section{Substitutional versus split-vacancy complexes with impurities}
\label{secV}

In this section we describe the results for a whole series of
calculations for impurity-vacancy complexes in Si and Ge. We would
like to find out for each of the 11 investigated impurities, which
complex, substitutional or split-vacancy, is most stable, and to
draw a border line between these two complex families as a function
of the ``size'' of the impurities. To achieve this aim, we have 
performed PPW calculations for vacancy complexes with the
impurities Al, Si, and P of the 3sp series, with Ga, Ge, As, and Se
of the 4sp series, with Cd, In, Sn, and Sb of the 5sp series and finally
with 6sp element Bi. In all cases only the neutral state has been
calculated.  First we discuss in detail the results for the Sn impurity in Ge. 

\begin{figure}[t]
\center{\includegraphics*[scale=0.29]{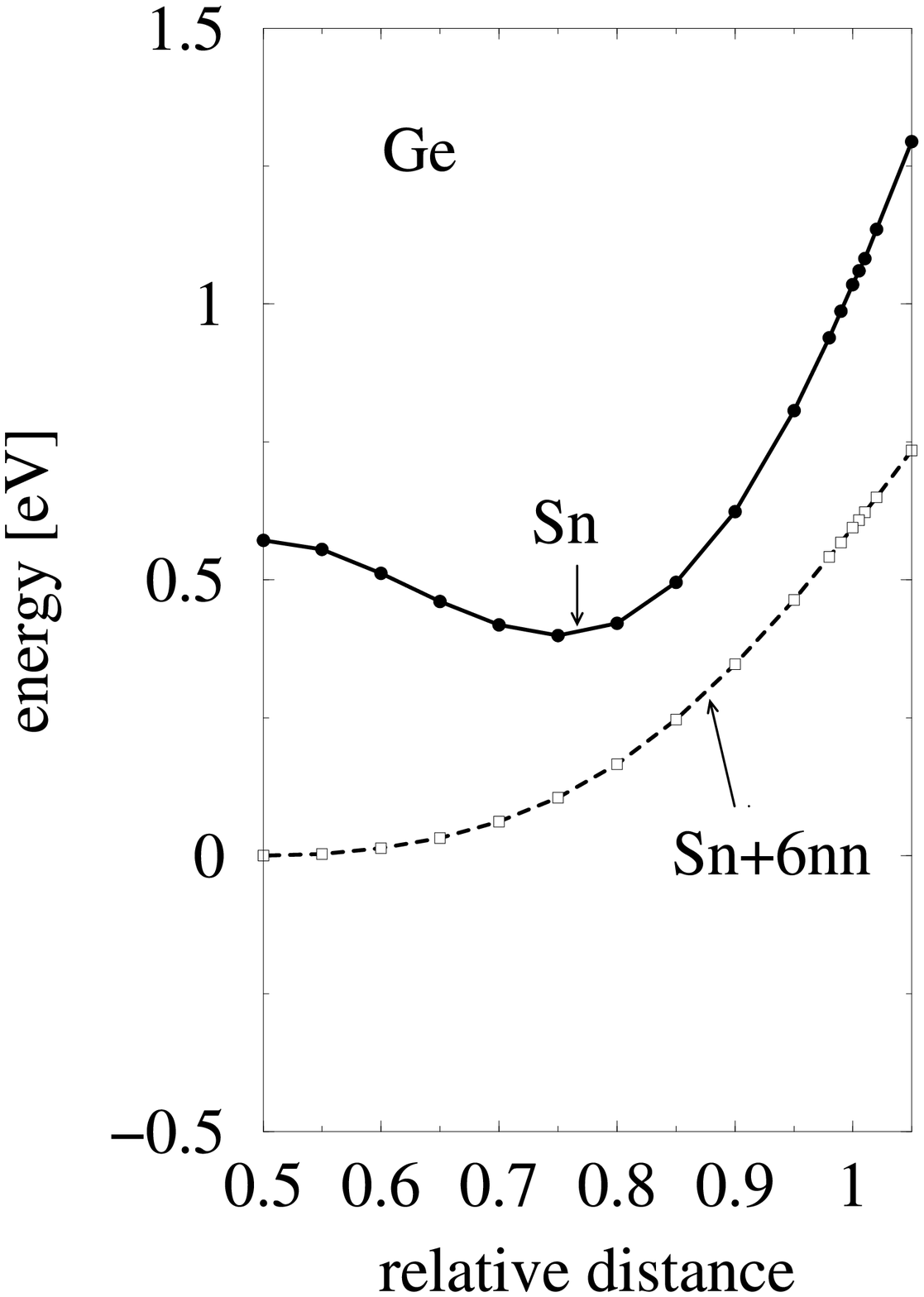}
\includegraphics*[scale=0.29]{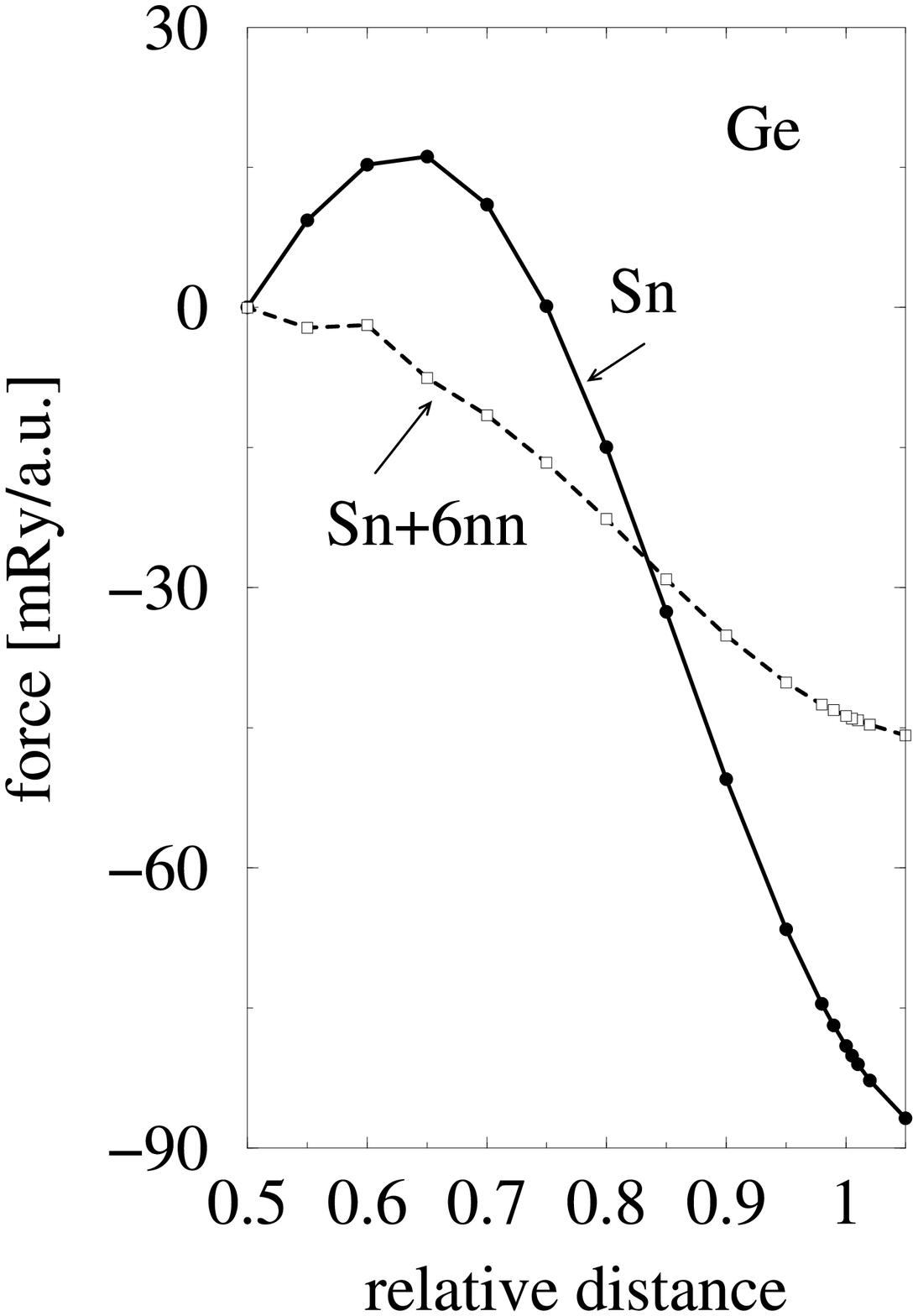}
\caption{Relaxations of the Sn-vacancy complex in Ge. 
Presented are PPW calculations with a cut-off energy of 13.67 Ry,
where only the Sn-atom is relaxed (full lines) and in addition the
six NN of the complex are relaxed (dashed lines). On the left, the total energy
is shown relative to the energy of the fully relaxed configuration with Sn
at the bond center site. On the right, the corresponding Hellmann-Feynman
forces on the Sn-Atom are shown.\label{fig:3}}} 
\end{figure}

Fig. \ref{fig:3} shows the total energy and the force on a Sn atom, if
this atom is moved adiabatically, i.e.\ by full relaxation of all other atoms,
from the substitutional position (indicated by ``relative distance 1'',
measured in units of the NN distance from the vacant site)
to the bond center (``relative distance 0.5'').
The two curves refer to the cases where only the Sn-impurity is allowed
to relax (full lines) and where in addition to the Sn atom also the six
host neighbors are relaxed (dashed lines). While by relaxation of only the 
Sn atom an intermediate position of about 0.75 is obtained, in the case
of full relaxations the Sn impurity moves without energy barrier into
the stable bond center position at 0.5.

The results for all 
considered 
impurities in both Si and Ge are listed in
Table \ref{over}. The second column gives the force on the impurity
in the unrelaxed substitutional position, i.e.\ when the impurity
and all neighbors are fixed at ideal lattice sites. The third column
gives the energy difference between both configurations, such that
for negative values the split configuration is stable and for positive
values the substitutional one. The last column gives the exact 
position of the impurity as measured in units of the NN distance
from the vacant site.

\begin{table}[t]
\center{
\begin{tabular}{lccc}
\hline
\hline
Impurity              &   force              & $\Delta E= E_{split}-E_{subst}$  & stable pos.  \\
    in Si                  &   [mRy] &      [eV]                    & in (111)  \\
\hline
Al   &  -22.4   &   -0.15 &    0.50 \\   
P   &   -11.5  &   +1.27 &    0.95 \\   
Ga   &  +5.37   &   +0.22 &    1.04 \\   
Ge   &   -21.7  & +0.30   &    0.87   \\     
As   &  -41.7   &  +0.82  &    0.90 \\      
Se   &    -44.9  &  +0.55  &   0.92 \\
${\mathrm{Cd}}^*$   & $\mbox{-85.2}^*$  &  $\mbox{-1.04}^*$  &   $\mbox{0.50}^*$  \\      
In   &  -77.2   &  -0.69  &    0.50  \\      
Sn   &   -105.  & -0.83   &  0.50  \\
Sb   &    -129.   & -0.68   &    0.50   \\     
Bi   &    -182.   & -0.92   &    0.50     \\  
\hline
\hline
Impurity              &   force              & $\Delta E= E_{split}-E_{subst}$  & stable pos.  \\
    in Ge                  &   [mRy] &      [eV]                    & in (111)  \\

\hline
Al   &  -0.96   &   +0.03 &    0.82 \\ 
Si   &  -4.5   &   +0.30 &    0.88 \\ 
P   &  -11.7   &   +0.81 &    0.93 \\ 
Ga   &  +14.6   &   +1.11 &    1.04 \\        
As   &  -44.1   &  +0.41  &    0.87 \\   
Se   &  -37.62   &  +0.35  &    0.91 \\   
${\mathrm{Cd}}^*$   & $\mbox{-69.6}^*$    & $\mbox{-1.01}^*$   &    $\mbox{0.50}^*$  \\      
In   &  -43.2   &  -0.47  &    0.50  \\         
Sn   &   -79.0  & -0.60   &  0.50  \\
Sb   &    -109.   & -0.64   &    0.50   \\     
Bi   &    -152.   & -0.86   &    0.50     \\  
\hline 
\hline                                                  
\end{tabular}
\caption{Relaxations of impurity-vacancy complexes in Si ($E_{cut}=$ 9 Ry, above) 
and Ge ($E_{cut}=$ 11.56 Ry, below):
The first column gives the forces on the impurity atom in the substitutional
positions, a negative sign means that forces are directed to towards the
the bond center. The second column gives the energy differences between
the fully relaxed configurations with the impurity at the bond-center and
at the substitutional site. The last column gives the final position of
the impurity in (111) direction; 0.5 is the bond center and 1.0 is
the substitutional lattice position. (* was obtained by the PAW method with 
$E_{cut}=$ 20.25 Ry).\label{over}}}
\end{table}

\begin{figure}[htb]
\center{\includegraphics*[scale=0.45]{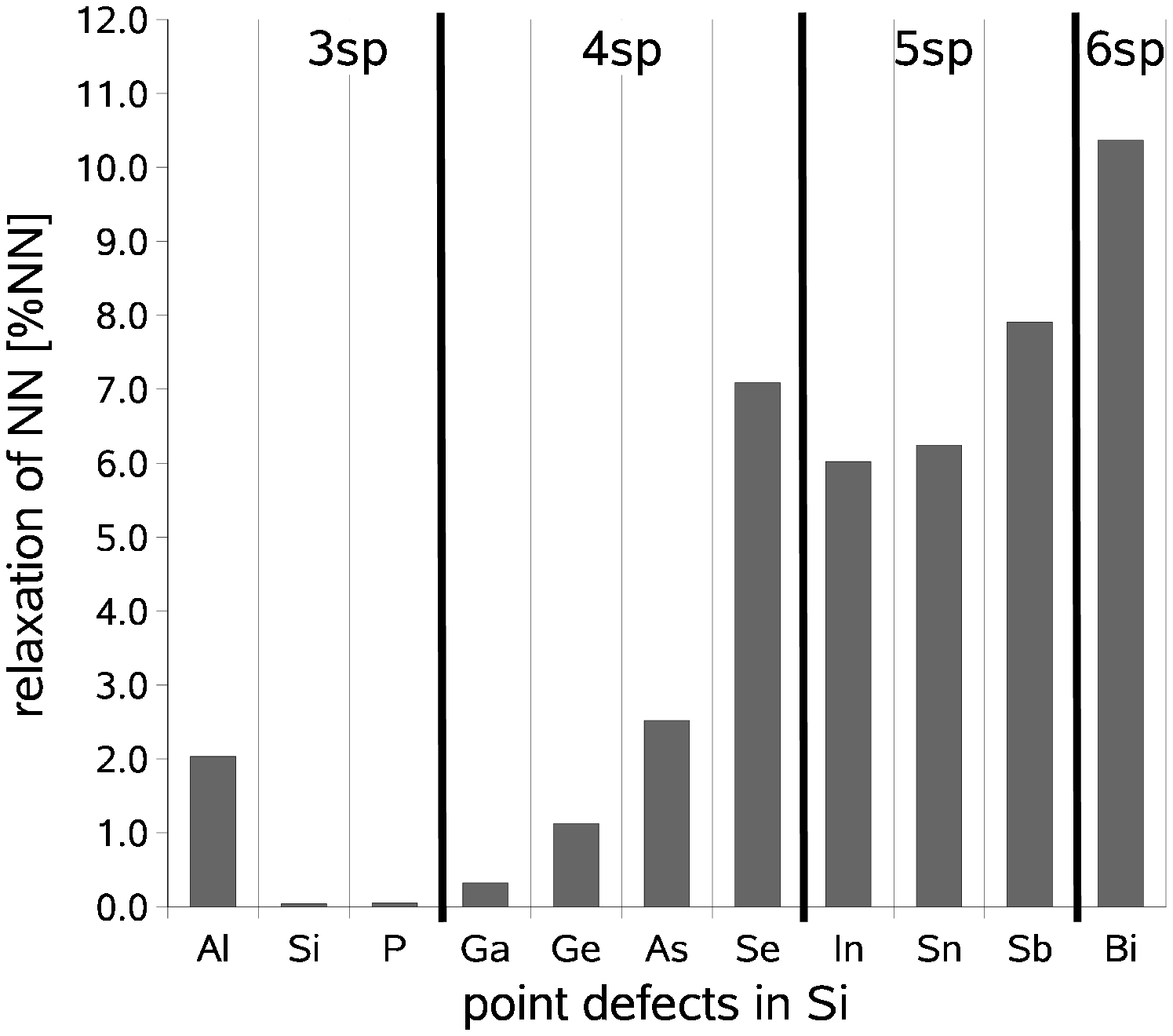}\\[0.25cm]
\includegraphics*[scale=0.45]{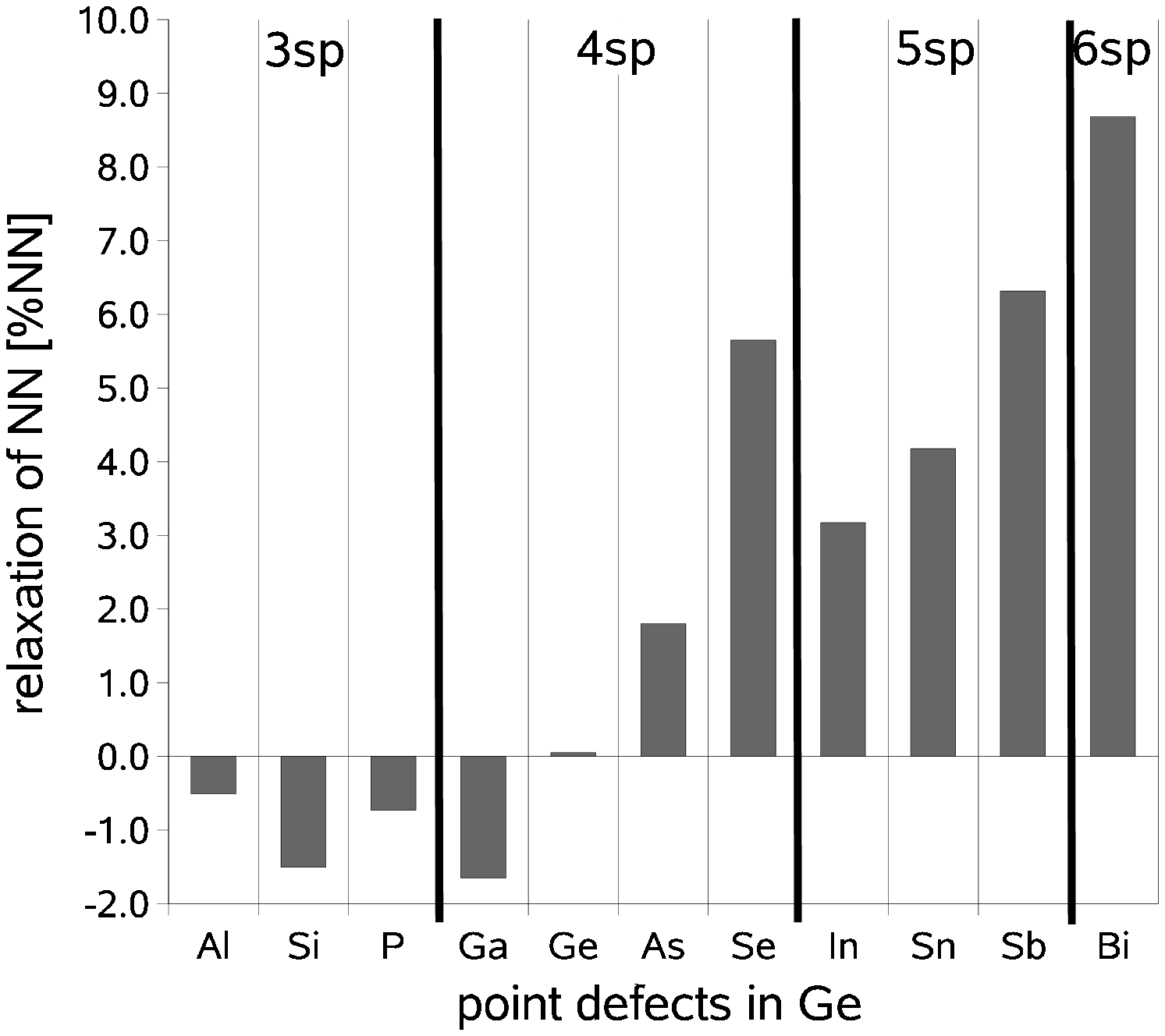}
\caption{Nearest-neighbor relaxations for various 3sp, 4sp, 5sp and 6sp impurities (substitutional point defects) in Si (above) and Ge (below). \label{fig:4}}} 
\end{figure}

The results show that both in Si and Ge the 5sp impurities Cd, In, Sn, and Sb
as well as the 6sp impurity Bi on the substitutional site experience
a strong force, which pushes them into the ``free space'' at the bond center,
thus gaining typically an energy of 0.5-1.0 eV. Note that the energy gains
are very substantial, much larger than the value 0.045 eV reported
in \cite{Kaukonen:00.1}. One would intuitively classify these heavy
impurities as oversized in Si and Ge. In line with this argument is the
fact that the forces as well as the resulting energy gains are somewhat
larger in Si than in Ge, due to the smaller lattice constant of Si. All
other impurities, with the exception of Al in Si, prefer the substitutional
configuration, however partly with large relaxations from the ideal
position. For instance, the Ge impurity in Si is shifted by 13\% of the 
NN distance in the direction of the vacancy and the As atom in Ge by
a similar amount. Of all impurities, only the Ga impurity in both Si
and Ge is shifted towards its remaining three neighbors, i.e.\ away
from the vacancy. Therefore it is tempting to classify the Ga impurity
as ``undersized''. The behavior of the Al impurity is most surprising.
As a member of the 3sp series it should be smaller than the isoelectronic
Ga impurity of the 4sp series. However the calculation seems to indicate
a larger size than Ga. The force at the ideal lattice site drives the
impurity in the direction of the vacancy, resulting in a split configuration
for the Al-vacancy complex in Si, and in a strongly distorted
substitutional configuration in Ge.

Unfortunately, we do not know any reliable rules for the ``size'' of 
impurities. Pauling \cite{Pauling:60.1} has given some general rules
about the volume of atoms, e.g., in a tetrahedral environment.
According to this, the size of impurities of the same row of 
the periodic table should decrease with increasing valence,
However, for the investigated cases we find just the opposite, i.e.\
the size increases with increasing valence, at least for the impurities 
of 4sp and 5sp series. 
For instance, As is larger than Ga, and Sb larger than In.
Moreover, according to Pauling \cite{Pauling:60.1}, the size of
isovalent impurities should increase with increasing main quantum number.
As expected, our calculations confirm this second rule in general.
However, deviations from both rules occur for Al: We find that Al is 
larger than Si and P, and in particular, is larger than the 
isovalent Ga impurity, which belongs to the higher 4sp series.

In order to shed some light on the size problem and its relation to
the stability of impurity-vacancy complexes, we have calculated the
lattice relaxations of the relevant isolated impurities in Si and Ge.
For simplicity, we have only calculated the displacements of the 
nearest neighbors, by fixing all other atoms at the ideal positions.
The idea behind these calculations is that both problems, i.e.\ the
NN relaxations of the impurities and the different configurations
of the impurity-vacancy complexes, are directly correlated. For instance,
the impurity exerts forces on the nearest neighbors, which, as a consequence,
are displaced from their ideal positions. The impurity itself is not
shifted, since the reaction forces of the nearest neighbors cancel
each other due to the tetrahedral symmetry. However, in the case of a
vacancy, one of the four neighbors is missing, so that the reaction forces
or the remaining three neighbors no longer cancel each other and shift
the impurity either to or away from the vacancy. In this way the pair
geometry should be related to the size of the single impurities.

The calculated lattice relaxations of the nearest neighbors of the 
various impurities are shown in Fig. \ref{fig:4}. They are given
in percentages of the NN distance, with positive values denoting
outwards relaxations. Note that in the calculations only the NN are
allowed to relax (the relaxation of the higher neighbor shells is
expected to increase the NN relaxations, presumable by about 30 to 50\%).
For the 4sp and 5sp impurities we clearly find that the relaxations
increase with increasing valence of the impurities. The same trend
has been found in Table \ref{over} for the force on the unrelaxed 
impurity. Also the relaxations of the 5sp impurities are in general
larger than the ones of the 4sp impurities, which is line with the fact
that the forces on the unrelaxed impurity in the substitutional
vacancy complex, as given in the second column of Table \ref{over},
show the same behavior. This makes it plausible that the 5sp and 6sp
impurities form a split-vacancy complex with the vacancy, while the 
4sp impurities prefer the substitutional complex. The figure also shows
the unusual behavior of the Al impurity, leading to a relatively
large outward relaxation in Si and only a small inward relaxation
in Ge. Thus the calculations clearly indicate that the size of Al is
larger than the one of Ga, which points to an unusual behavior of Al,
showing also up in the stability of the Al-split-vacancy complex
in Si-bulk. Thus the 
NN relaxations of the isolated impurities correlate qualitatively well
with the stability of the impurity-vacancy complexes. However a detailed
comparison is not possible. For instance, the outward NN relaxations
of the Al impurity in Si are slightly smaller than the ones of As;
yet Al forms with the vacancy in Si a split complex, but As a 
substitutional one. Similarly, the NN relaxations of the Si impurity
in Ge are inwards, but in the substitutional Si-vacancy complex Si
relaxes by 12\% towards the vacancy.

\section{Summary}
\label{secVI}

In this paper we have studied the electronic and geometrical structure
of vacancy complexes with ``oversized'' and ``normal-sized''
impurities in Si and Ge. First we discuss the electronic structure
of the Sn-vacancy complex. In agreement with recent {\it{ab initio}} studies
for the Sn-vacancy complex in Si \cite{Kaukonen:00.1,Kaukonen:01.1}
and in analogy to our recent studies \cite{Hoehler:04.1} for the Cd-vacancy
complex in Si and Ge, we find
a split-vacancy complex as the stable configuration in both hosts
with the impurity on the bond center position
and the vacancy split into two ``half vacancies'' on the neighboring
sites. The density of states of Sn and Cd are characterized by the same
sequence of states and similar charge states exist. The calculated
hyperfine fields of Sn and the Si neighbors are in good agreement with
the measurements of Watkins \cite{Watkins:75.1}; however, no
agreement is obtained with the available isomer shift data for
Sn\cite{Weyer:80.1,Weyer:80.2}.

In the second part of the paper we present a systematic study of vacancy
complexes with 11 different impurities, both in Si and Ge. We find that, in
both hosts, impurities of the 5sp and 6sp series form split-vacancy
complexes, while impurities of the 4sp and 3sp series prefer,
more or less distorted, substitutional complexes. An exception from
this rule is Al, forming in Si a split complex and in Ge a strongly
distorted substitutional complex. Qualitatively we explain the results in
terms of the size of these impurities, such that oversized impurities
can lower their energy in the ``free space'' available at the bond
center site. To examine the ``size'' of these impurities, we calculated
the nearest neighbor relaxations of the isolated impurities and find
a good correlation between the calculated NN relaxations and the 
structure of the impurity-vacancy complexes.

\begin{acknowledgments}
We thank R. Sielemann for helpful and motivating discussions.
We gratefully acknowledge financial support by the German government,
BMBF-Verbundforschung, project 05KK1CJA/2.
\end{acknowledgments}

\end{document}